\documentclass[12pt]{iopart}  
\usepackage[dvips]{graphicx}   

\def \lta {\mathrel{\vcenter
     {\hbox{$<$}\nointerlineskip\hbox{$\sim$}}}}
\def \gta {\mathrel{\vcenter
     {\hbox{$>$}\nointerlineskip\hbox{$\sim$}}}}

\def\beqra{\begin{eqnarray}}
\def\eeqra{\end{eqnarray}}
\def\beq{\begin{equation}}
\def\eeq{\end{equation}}

\def\a{\alpha}
\def\BF{B_F(\phi)}

\def\D{\Delta}

\begin{document}            
\begin{flushright}
DFPD--05/TH/07
\end{flushright}

\title{Cosmological evolution of Alpha  driven by a general coupling with Quintessence}  
\author{Valerio Marra, Francesca Rosati\footnote{Contact e-mail: francesca.rosati@pd.infn.it, valerio.marra@pd.infn.it}}     
\address{Dipartimento di Fisica ``Galileo Galilei'' and INFN \\
Universit\`a di Padova, via Marzolo 8, I-35126 Padova, Italy}            
\begin{abstract}
A general model for the cosmological evolution of the fine structure constant $\a$ driven by a typical Quintessence scenario is presented.  
We consider a coupling  between the Quintessence scalar $\phi$ and  the electromagnetic kinetic term $F_{\mu\nu}F^{\mu\nu}$, given by a general function $B_F(\phi)$. 
We study the dependence of the cosmological $\D\a(t)$ upon the functional form of $\BF$ and discuss the constraints imposed by the data.
We find that different cosmological histories for $\D\a(t)$ are possible within the avaliable constraints. We also find that Quasar absorption spectra evidence for a time variation of $\a$, if confirmed, is not incompatible with Oklo and meteorites limits.
\end{abstract}
\maketitle    
\newpage

%
\section{Introduction}      
Over the last few years there has been increasing interest in the possibility of varying the fundamental constants over cosmological time-scales. This has a twofold motivation. 
On one side, several observations point towards the existence of a smooth dark energy component in the universe, which could be modeled via a dynamical scalar field (Quintessence) \cite{quint}. In general, we expect such a cosmological scalar to couple with some, if not all, the terms in the matter-radiation Lagrangian, thus inducing a time variation of physical masses and couplings.
On the other side, recent improved measurements on possible variations of the fundamental constants are opening up the possibility of testing the theoretical models to a good degree of precision over a wide range of cosmological epochs. It should also be mentioned that, although controversial, some evidence of time variation of the fine structure constant $\a$ in Quasar absorption spectra was recently reported \cite{Webb,Murphy}.
The cosmological variation of fundamental constants induced by couplings with the Quintessence scalar is then worth studying  in order to see if such a field could be responsible for a measurable effect.

Among all the possibilities, the time-variation of the fine-structure constant is the simplest to study both from the theoretical and experimental points of view. In this paper we will then restrict ourselves to this issue.
The theoretical study of a time-varying fine structure constant dates back to 1982 when Beckenstein \cite{Bekenstein:1982eu} first considered the possibility of introducing a linear coupling between a scalar field and the electromagnetic field. More recently the Beckenstein model has been revived, generalized and confronted with updated experimental limits 
\cite{Carroll:1998zi,Olive:2001vz,Damour:2002nv,alpha-models}.
The concrete case of the Quintessence scalar  has been considered too 
\cite{Copeland:2003cv,Lee:2004vm,alpha-quint1,alpha-quint2}. However, as we will see, most authors restrict their study to the simplest case of a linear  or quadratic coupling. The possibility of reconstructing the dark energy equation of state from a measure of $\a$-variation has also been proposed in the literature \cite{alpha-w}.

In this paper we will discuss a general model for the variation of the fine structure constant $\a$ driven by a typical Quintessence scenario. After briefly reviewing the most recent observational and experimental constraints on the variation of $\a$, we will go on to construct the theoretical framework. 
In particular we will consider the case of a general coupling $\BF$ (see Eq.~(\ref{BF}) below) which includes several classes of possible functions. In this way we will be able to study the dependence of the cosmological variation of alpha, $\D\a(t)$, upon the functional form of $\BF$ and discuss the constraints imposed by present data.
We will find that different cosmological histories for $\D\a(t)$ are possible within the avaliable constraints. We will also find that, Webb et al.~data \cite{Webb,Murphy}, if confirmed, are not incompatible with Oklo and meteorites constraints \cite{Olive:2002tz,Damour:1996zw}.

\section{Overview of the constraints}
\begin{figure}[!htb]
\begin{flushright}
\includegraphics[width=14cm]{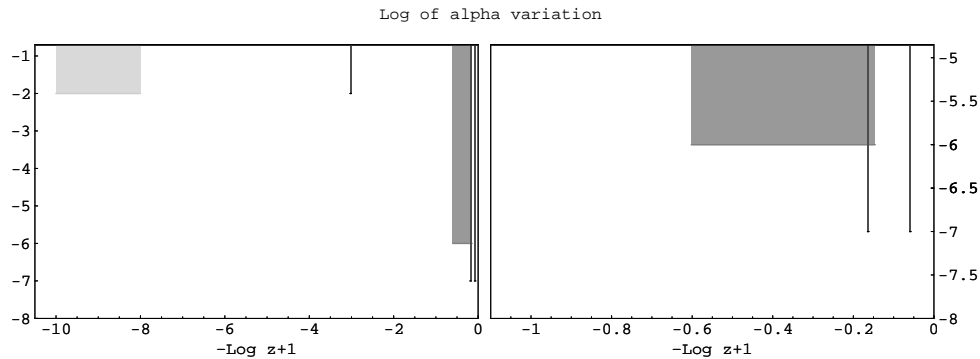}
\caption{The experimental constraints (1)-(6) discussed above are summarized in the picture: $\log | \Delta \alpha / \alpha |$ is plotted as a function of the redshift $z$. 
On the right-hand side we zoom on $z\lta 10$. The grey areas are those  excluded by present data.}
\label{limiti}
\end{flushright}
\end{figure}

Comprehensive reviews about the theoretical and experimental issues connected to the time variation of fundamental constants can be found in Refs.~\cite{Uzan:2002vq}, \cite{Olive:2002tz} and \cite{Martins:2004ni}. In the following we summarize the avaliable constraints on  the time variation of $\alpha$, expressed as functions of the redshift $z$ (see also Fig.~\ref{limiti}):
\begin{equation}
{\Delta \alpha(z) \over \alpha}\equiv {\alpha(z) - \alpha_{0}\over \alpha_{0}}
\end{equation}
where $\alpha_{0}=\alpha(0)$ is the value measured today. 

(1) The most ancient data come from Big Bang Nuclesynthesis (BBN) and give \cite{bbn,bbncmb}:
\begin{equation}
\left | {\Delta \alpha \over \alpha}\right | \lta 10^{-2}
\qquad \qquad z=10^{10}-10^{8}
\; \;  .
\end{equation}

(2) More recently we have the limit coming from the power spectrum of anisotropies in the Cosmic Microwave Background (CMB)
 \cite{bbncmb}:
\begin{equation}
\left | {\Delta \alpha \over \alpha}\right | < 10^{-2}
\qquad \qquad z=10^{3}
\;\; .
\end{equation}

(3) From absorption spectra of distant Quasars there are  more controversial data.
Webb and Murphy's groups combined data \cite{Webb,Murphy} report  a $4 \, \sigma$ evidence for $\a$ variation: $\Delta \alpha / \alpha=(-0.543 \pm 0.116) \cdot 10^{-5}$ on a cosmological time span between $z=0.2$ and $z=3.7$. This result has not been confirmed by other groups. For example, Chand et al.~\cite{Chand} give: ${\Delta \alpha / \alpha} =(-0.06 \pm 0.06) \cdot 10^{-5}$ for $z=2.3-0.4$ and
${\Delta \alpha / \alpha} =(0.15 \pm 0.43) \cdot 10^{-5}$ for $z=2.92-1.59$. 
We have chosen to be conservative and will consider a limit based on the last two results, which is consistent with zero variation:
\begin{equation} \label{quasa}
\left |{\Delta \alpha \over \alpha} \right | \lta 10^{-6}
\qquad \qquad z=3-0.4
\;\;  .
\end{equation}

(4) From the analysis of the ratio Re/Os in meteorites dating around 4.56 billion years ago it is possible to compute $^{187}$Re half-life, which gives \cite{Olive:2002tz}:
\begin{equation}
\left | {\Delta \alpha \over \alpha}\right | \lta 10^{-7}
\qquad \qquad z=0.45
\;\;  .
\end{equation}

(5) From the Oklo natural nuclear reactor that operated 2 billion years ago in Gabon, we also have \cite{Olive:2002tz,Damour:1996zw}:
\begin{equation}
\left | {\Delta \alpha \over \alpha}\right | \lta 10^{-7}
\qquad \qquad z=0.14
\;\; .
\end{equation}

(6) We then have limits coming from laboratory measurments which constrain the present rate of change of $\a$.
Comparing atomic clocks, which use different transitions and atoms, what is obtained is \cite{Marion:2002iw}:
\begin{equation} \label{labo}
\left | {\dot{\alpha} \over \alpha}\right | \lta 10^{-15} \mbox{ yr}^{-1}
 \qquad \;\; z=0
\end{equation}
where the dot represents differentiation w.r.t. cosmic time.

%
%

(7) In addition to the limits discussed above, there is a constraint coming from indirect violation of the Weak Equivalence Principle (WEP), parametrized by the  E\"otv\"os ratio
\begin{equation}
\eta=2{|a_{1}-a_{2}| \over |a_{1}+a_{2}|}
\end{equation}
where $a_{1}$ e $a_{2}$ are the accelerations of two different test bodies in the Earth gravitational field. 
These constraints come from the fact that nucleon masses get electromagnetic corrections from quark-quark interactions. As extensively discussed in \cite{Gasser:1982ap}, the leading term of the electromagnetic contribution comes from the electrostatic energy of the quark distribution, which is proportional to $\alpha$.
The corrected masses, to leading order in $\alpha$,  can then be written as \cite{alpha-quint1,Gasser:1982ap}:
\begin{eqnarray} \label{gama}
m_{p}&=&m+\alpha \: B_{p}
\nonumber\\
m_{n}&=&m+\alpha \: B_{n}
\end{eqnarray}
where $p,n$ stand for proton and neutron and $B_p \equiv 0.63 \mbox{MeV}/\a_0$,  $B_n \equiv -0.13 \mbox{MeV}/\a_0$.
If we suppose that $\a=\a(\phi)$, then we will induce a $\phi$-dependence on the nucleon masses:
\begin{equation} \label{me}
\delta m_{n}=B_n   \delta \alpha    \qquad  ;  \qquad  \delta m_{p}=B_p  \delta \alpha \;\;  .
\end{equation}
If we define
\beq
g_{i} =
{\partial m_{i} \over \partial \phi} =
{\partial \alpha \over \partial \phi} B_i
\label{gi}
\eeq
we get an indirect violation of the equivalence principle induced by the `fifth-force' mediated by the scalar field
\begin{equation} \label{eta}
\eta \simeq {M_{Pl}^{2} \over 4 \pi \bar{m}^{2}} \left ( R_{n}^{E} g_{n}+R_{p}^{E} g_{p} \right ) \left ( \Delta R_{n} g_{n}+\Delta R_{p} g_{p} \right )
\end{equation}
where:
\begin{equation}
R_{i}^{E}\equiv {n^{E}_{i} \over n^{E}_{n}+n^{E}_{p}}\simeq 0.5
\qquad
\Delta R_{i}\equiv {|n_{i,\: 1}-n_{i,\: 2}| \over n_{n}+n_{p}}\simeq 0.06-0.1 \,\,\, ,
\end{equation} 
and $\bar{m} \simeq 931~ \mbox{MeV}$ is the atomic mass unit. The suffix $E$ refers to the Earth, while $1$ and $2$ refer to two test bodies having equal mass but different composition.
From Eqs.~(\ref{gi})-(\ref{eta}) we see that any model of $\a$-variation will induce a characteristic $g_{p,n} \not = 0$ and hence WEP violation:  while the first two factors in Eq.~(\ref{eta}) are universal and depend on the Earth composition, the third term is not zero if and only if $g_{p,n}\not=0$ and the test bodies have different composition in neutrons and protons. The current limits on WEP violations impose \cite{equiv}:

\begin{equation} \label{eot}
\eta < 10^{-13}  \,\,\, .
\end{equation}

\section{The theoretical framework}
Following Olive et al.~\cite{Olive:2001vz}, the most generic action involving a scalar field, the Standard Model fields and an hypothetical Dark Matter particle $\chi$, can be written as
\beqra  \label{action}
S & = & \frac{1}{16 \pi G} \int d^4x \sqrt{-g}~R + 
\int d^4x \sqrt{-g} \left[   \frac{1}{2} \partial^{\mu}\phi\partial_{\mu}\phi - V(\phi) \right]
\nonumber \\
& - & {1 \over 4} \int d^4x \sqrt{-g} \ B_{F}(\phi)F_{\mu \nu}F^{\mu \nu}  
-  {1 \over 4} \int d^4x \sqrt{-g} \  B_{F_{i}}(\phi)F^{(i)}_{\mu \nu}F^{(i)\mu \nu} 
\nonumber  \\
& + & \int d^4x \sqrt{-g} \  \sum_{j} \left[ \bar{\psi}_{j} D\!\!\!\!/ \psi_{j}+i B_{j}(\phi)m_{j}\bar{\psi}_{j}\psi_{j} \right] 
\nonumber  \\
& + &  \int d^4x \sqrt{-g} \  \left[ \bar{\chi} \partial\!\!\!/ \chi - B_\chi (\phi) m_{\chi} \chi^T \chi  \right]
\eeqra
where $D\!\!\!\!/=\gamma_{\mu}D^{\mu}$ and for the electromagnetic term, for example, $D^{\mu}=\partial_{\mu}-i e_{0} A_{\mu}$. The index $i=1,2,3$ refers to the $SU(3)$ gauge group of the Standard Model and  $j$ runs over the various matter fields.

The form of the action (\ref{action}) follows from supplying $\phi$-dependent factors to all mass and kinetic terms to the standard Lagrangian (which would have all $B_i=1$).
In general we would expect that all of the $B_i(\phi)$'s are switched on, if not forbidden by any symmetry principle. However, the theoretical treatment of the full Lagrangian is very cumbersome and so the coupling functions $B_i(\phi)$ are usually switched on one at a time. In this way one can also disentangle the effects due to each single term. 
In this paper we want to focus on the fine-structure constant $\a$, and so we will keep only $B_F(\phi)\not = 1$ and set all the other functions equal to 1. 

The relevant part of the action for the effect we are going to study is then
\beqra  \label{action2}
S  = \frac{1}{16 \pi G} \int d^4x \sqrt{-g}~R + 
\int d^4x \sqrt{-g} \left[   \frac{1}{2} \partial^{\mu}\phi\partial_{\mu}\phi - V(\phi) \right]
\nonumber \\
 -  {1 \over 4} \int d^4x \sqrt{-g} \ B_{F}(\phi)F_{\mu \nu}F^{\mu \nu}  
\eeqra
which allows to define an ``effective'' fine structure constant
\begin{equation} \label{alpha}
\alpha(t)={\alpha_{0}\over B_F(\phi(t))}
\end{equation}
where $\alpha_{0}$ is the value measured today. From (\ref{alpha}) we obtain the relative variation relevant for each cosmological epoch
\begin{equation} \label{Dalpha}
{\Delta \alpha \over \alpha}\equiv {\alpha(t) - \alpha_{0}\over \alpha_{0}}={1-B_F(\phi(t))\over B_F(\phi(t))}
\end{equation}
It can immediately be seen that, depending on the cosmological evolution of $\phi(t)$ and on the functional form of $B_F(\phi)$, the fine structure constant $\a$ could in principle have had many possible histories during the life-time of the universe. 
What possibilities are allowed by a general coupling $\BF$ within the avaliable observational constraints is then worth studying. 

%
%
The relevant equations governing the cosmological evolution in a flat universe are the following
\beqra
\frac{\ddot{a}}{a} = - \frac{4 \pi}{3 M_p^2} ~\sum_i (1+3 w_i)\rho_i
\label{ddota} \\
H^2 \equiv \left( \frac{\dot{a}}{a} \right)^2 = \frac{8\pi}{3M_p^2}~ \sum \rho_i 
\label{H2}\\
\ddot{\phi} + 3H\dot{\phi} + \frac{dV}{d\phi} = 0
\label{phi-eq}
\eeqra
where $i=m,~r,~\phi$ runs over the matter (including dark matter), radiation and scalar components. The relevant equations of state are $w_m=0$ for matter, $w_r=1/3$ for radiation and $w_\phi$ as defined in Eq.~(\ref{wphi}).
It is important to note that the evolution equation of the Quintessence scalar (\ref{phi-eq}) does not depend on $B_F$ or its derivatives. This is due to the fact that the statistical average of the term $F^{\mu\nu}F_{\mu\nu}$ over a current state of the universe is zero. So the only term that drives $\phi$ during the cosmological evolution is the potential $V(\phi)$.

Since we are working under the hypothesis that the scalar field $\phi$ in Eq.~(\ref{phi-eq}) is the Quintessence scalar, we should also impose the additional constraints coming from Quintessence phenomenology.
In particular we choose a runaway potential which goes to zero as far as the field $\phi$ rolls to infinity, in accordance with the observational data. It is also required that the scalar dynamics gives the correct value for the equation of state 
\begin{equation}
\label{wphi}
w_\phi = \frac{\dot{\phi}^2/2 - V(\phi)}{\dot{\phi}^2/2 + V(\phi)} \;\;\; \left( \lta -0.7 \; \mbox{today} \right)
\end{equation}
The most general form for the Quintessence potential involves a combination of a power-law and exponential terms \cite{Ng:2001hs}. For the purpose of this paper, however, we will consider the simplest case of an inverse power-law potential $V(\phi)=M^{4+n}\phi^{-n}$, which gives a late-time attractor equation of state $w_\phi=-2/(n+2)$ during matter domination \cite{Steinhardt:1999nw}. The potential should also be normalized  in  order to give the correct energy density today
($\rho_\phi^0 \simeq V(\phi) \simeq 2/3~ \rho_c^0$): this sets the mass scale $M$. In what follows we will choose $n=1$ in the potential in order to have the correct attractor equation of state, and so obtain $M\simeq\sqrt[5]{2/3~\rho_c^0 ~M_p}$.

We have checked that choosing different Quintessence potentials gives a subdominant effect on the cosmological variation of $\a$, with respect to changing the coupling function $\BF$.
In what follows we will then fix $V(\phi)=M^5/\phi$ and study the effect of different $\BF$'s.
An interesting study, which is complementary to what is done here, is that of Ref.~\cite{Copeland:2003cv} where the effect induced by different Quintessence models on the cosmological $\D\a$ is examined in detail, while keeping the function $\BF$ fixed.

%
%

In order to be as general as possible we will consider a function $\BF$ which is a combination of different possible behaviors and characterized by a set of four parameters that are allowed to vary freely:
\beq
\label{BF}
B_F(\phi) = \left(\frac{\phi}{\phi_{0}}\right)^{\epsilon}  
\left[1-\zeta {(\phi-\phi_{0})}^{q} \right] \ e^{\tau(\phi-\phi_0)}   \,\,\,\, .
\eeq
This choice is not motivated by a specific theoretical model, but is rather a working tool for  obtaining different functional forms of $\BF$ and thus cosmological histories of $\a$, according to Eq.~(\ref{Dalpha}).
We have chosen a combination of functions (power-law, polynomial, exponential) that can be switched on and off at will (depending on the values of the parameters $\epsilon$, $\zeta$, $\tau$ and $q$), thus giving rise to a variety of possibile $\BF$'s. In this way we can carry on a unified discussion of a number of different models of $\a$ variation.

\begin{figure}[!htbp]
\begin{flushright}
\includegraphics[width=14cm]{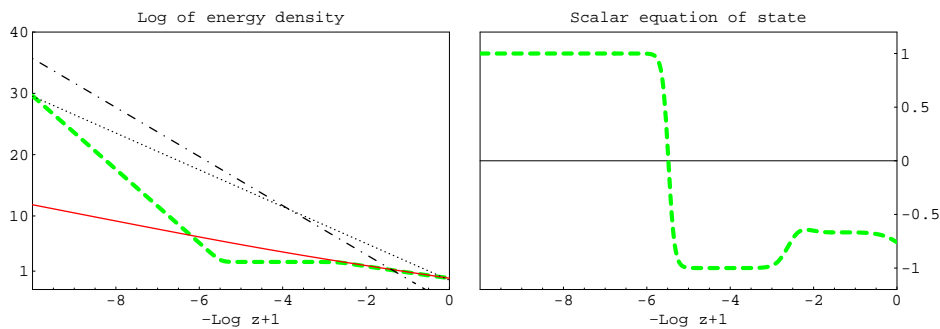}
\caption{Evolution of the energy densities (left) and scalar equation of state (right) for a  quintessence model with potential $V=1/\phi$ and  initial conditions $\rho_{\phi}^{in}/\rho_{c}^{0}=10^{30}$ at $z=10^{10}$.
The dot-dashed line represents the energy density of radiation, the dotted line the energy density of matter, the green dashed line the energy density of quintessence and the red solid line the attractor. All of the energy densities are expressed in units of the present critical energy density  $\rho_{c}^{0}$.}
\label{quinti}
\end{flushright}
\end{figure}

%
\section{Cosmic evolution of $\alpha$}
We have numerically solved the cosmological equations (\ref{ddota})-(\ref{phi-eq}) and then plotted the resulting cosmological history of $\D\a$ for various classes of functional forms of $\BF$, according to Eq.~(\ref{Dalpha}).
As already mentioned, for illustrative purposes we have chosen a scalar potential $V=1/\phi$ and initial conditions $\rho_{\phi}^{in}/\rho_{c}^{0}=10^{30}$ at $z=10^{10}$.
Fig.~\ref{quinti} shows the corresponding evolution of the energy densities and of the scalar equation of state parameter.

%
\subsection*{Linear coupling}

The simplest case is given by the choice $\epsilon=\tau=0$ and $q=1$ for the parameters in Eq.~(\ref{BF}):
\begin{equation} \label{beki}
B_{F}(\phi) = 1-\zeta  (\phi-\phi_{0})
\end{equation}
This case corresponds to the original Beckenstein proposal \cite{Bekenstein:1982eu}, which however did not supply a potential to the scalar field\footnote{To be precise,  Beckenstein actually invoked an exponential coupling, which however is practically equivalent to  eq.(\ref{beki}) due to the smallness of $\zeta\phi$.}. 
Copeland et al.~\cite{Copeland:2003cv} give a comprehensive discussion on various Quintessence models linearly coupled to the electromagnetic field, but assuming  Webb et al.~data \cite{Webb,Murphy} to be correct and imposing on $\D\a(t)$ to agree with that measure.

\begin{figure}[!htbp]
\begin{flushright}
\includegraphics[width=14cm]{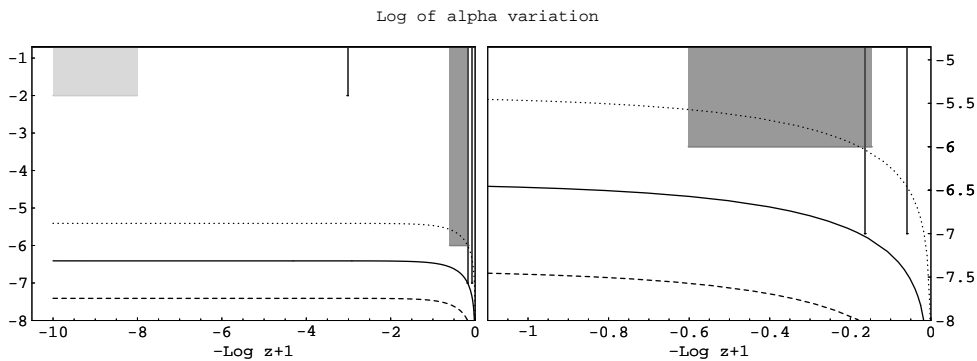}
\caption{The logarithm of $| \Delta \alpha / \alpha |$ is plotted as a function of Log$(z+1)$ for $B_{F}(\phi) = 1-\zeta  (\phi-\phi_{0})$ with $\zeta=0.6 \cdot 10^{-5}$ (dotted line), $\zeta=0.6 \cdot 10^{-6}$ (solid line) and $\zeta=0.6 \cdot 10^{-7}$ (dashed line).  
On the right-hand side we zoom on $z\lta 10$. 
Only the curves not overlapping the grey areas are phenomenologically viable.}
\label{linear}
\end{flushright}
\end{figure}

In our case, the resulting $\Delta \alpha$, as defined in Eq.~(\ref{Dalpha}), is plotted in Fig.~\ref{linear}. We tried a number of different values for $\zeta$, in order to verify in which cases all the available experimental constraints were simultaneously satisfied. We found that they are all respected for $\zeta \leq0.6 \cdot 10^{-6}$.
With this choice, the constraints on the violation of equivalence principle and the constraints derived from atomic clocks are automatically satisfied:
\begin{equation}
\eta \simeq 4 \cdot 10^{-21} \ll 10^{-13}
\qquad \qquad
\left | {\dot{\alpha} \over \alpha_{0}}\right |
=4 \cdot 10^{-17}
\ll 10^{-15} \mbox{ yr}^{-1}
\end{equation}
%

%
%

\subsection*{Polynomial coupling}

A slightly more complicated case is given by the choice  $\epsilon=\tau=0$, allowing the exponent $q$ to be $>1$:
\begin{equation} \label{beki2}
B_{F}(\phi)= 1-\zeta  (\phi-\phi_{0})^{q}
\end{equation}
The case  of a quadratic coupling ($q=2$) was considered in Ref.~\cite{Lee:2004vm}, but with the additional assumption of a proportionality relation between $\BF$ and $V(\phi)$.

\begin{figure}[!thbp]
\begin{flushright}
\includegraphics[width=14cm]{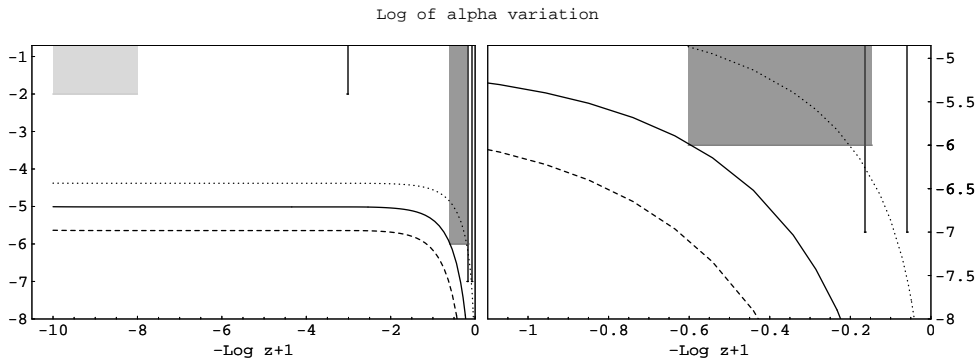}
\caption{The logarithm of $| \Delta \alpha / \alpha |$ is plotted as a function of Log$(z+1)$ for $B_{F}(\phi) = 1-\zeta  (\phi-\phi_{0})^{q}$ with $\zeta = 10^{-4}$ and $q=3$ (dotted line), $q=6$ (solid line) and $q=9$ (dashed line). 
On the right-hand side we zoom on $z\lta 10$.  
Only the curves not overlapping the grey areas are phenomenologically viable.}
\label{polinomial}
\end{flushright}
\end{figure}

We have found  that the data do not impose any upper limit on the exponent $q$ and that increasing $q$ makes it possible to reduce the fine-tuning in $\zeta$.
For example, choosing $\zeta = 10^{-4}$ the experimental limits are respected for $q=6$ and  the constraints on the violation of equivalence principle and the constraints derived from atomic clocks satisfied by many orders of magnitude.
In Fig.~\ref{polinomial} we plot Log~$|\Delta \alpha / \alpha |$ for $\zeta = 10^{-4}$ with $q=3, \, 6 \mbox{ and } 9$, as function of red-shift.
%
%

\begin{figure}[!thbp]
\begin{flushright}
\includegraphics[width=14cm]{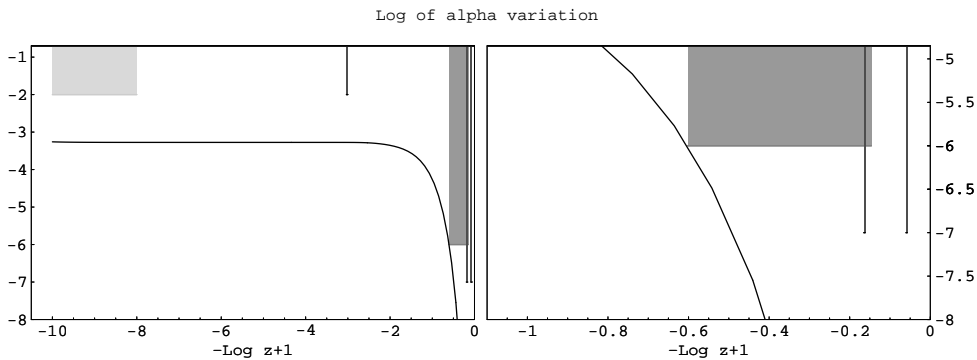}
\caption{The logarithm of $| \Delta \alpha / \alpha |$ is plotted as a function of Log$(z+1)$ for $B_{F}(\phi) = 1-\zeta  (\phi-\phi_{0})^{q}$ with $\zeta = 1$ and $q=17$  (solid line). 
On the right-hand side we zoom on $z\lta 10$.  Note that all the experimental limits  are satisfied without any fine--tuning in the parameters of the function $\BF$.}
\label{polyb}
\end{flushright}
\end{figure}

As already mentioned, increasing the exponent $q$ we can do even better.
For example, with $q=17$ the experimental constraints are satisfied even for $\zeta=1$, as illustrated in Fig.~\ref{polyb}. It should be emphasized that among all the possibilities considered in this paper, this choice of the parameters appears to be the most natural of all. A notable feature is also the fact that the value of $\D\a$ is enhanced in the past, with respect to the other cases, becoming closer to the observational limits, while falling off very steeply in recent times.
%
%

\subsection*{Power--law coupling}
\begin{figure}[htbp]
\begin{flushright}
\includegraphics[width=14cm]{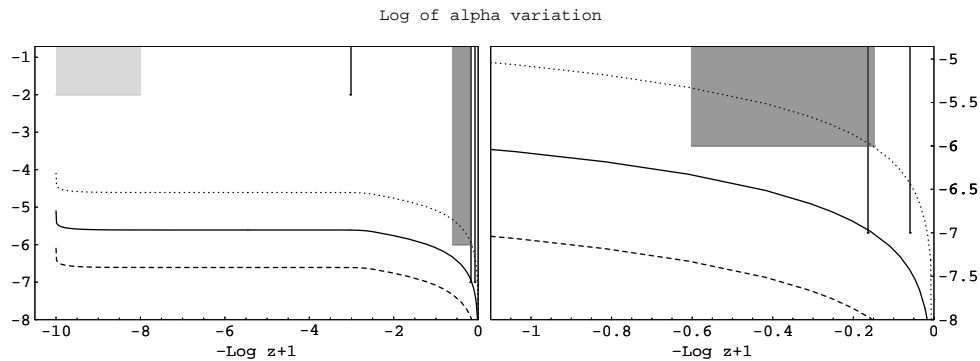}
\caption{The logarithm of $| \Delta \alpha / \alpha |$ is plotted as a function of Log$(z+1)$  for $B_{F}(\phi)=\left({\phi \over \phi_{0}}\right)^{\epsilon}$ with $\epsilon=4 \cdot 10^{-6}$ (dotted line), $\epsilon=4 \cdot 10^{-7}$ (solid line) and $\epsilon=4 \cdot 10^{-8}$ (dashed line). 
On the right-hand side we zoom on $z\lta 10$. 
Only the curves not overlapping the grey areas are phenomenologically viable.}
\label{power}
\end{flushright}
\end{figure}

With the choice $\zeta=\tau=0$ we obtain the following coupling function:
\begin{equation} \label{beki5}
B_{F}(\phi)=\left({\phi \over \phi_{0}}\right)^{\epsilon}  \,\, .
\end{equation}
In this case, it is necessary to fine-tune the exponent $\epsilon$ in order to satisfy the data, due to the smallness of $\phi$ in the early universe. Keeping $\epsilon$ of order one would violate even the constraints from BBN.
We found that the experimental limits are respected for  $|\epsilon| \leq 4 \cdot 10^{-7}$.
In Fig.~\ref{power} we plot Log~$|\Delta \alpha / \alpha |$  as a function of red-shift for different choices of $\epsilon$. Note that the sign of $\Delta \alpha$ depends on the sign of $\epsilon$.
With the choice $\epsilon=4 \cdot 10^{-7}$, the constraints on the violation of equivalence principle and the constraints derived from atomic clocks are automatically satisfied:
\begin{equation}
\eta \simeq 4 \cdot 10^{-21} \ll 10^{-13}
\qquad \qquad
\left | {\dot{\alpha} \over \alpha_{0}}\right |
=4 \cdot 10^{-17}
\ll 10^{-15} \mbox{ yr}^{-1}
\end{equation}
Such a small exponent might look quite unnatural, however Eq.~(\ref{beki5}) for $\epsilon \ll 1$ is equivalent to:
\begin{equation} \label{bk5}
B_{F}(\phi)=1+\epsilon \ln \left({\phi \over \phi_{0}}\right)    \,\, .
\end{equation}
In this way the fine tuning is moved from the exponent to the coefficient.
%
%

\subsection*{Exponential coupling}

The choice $\epsilon=\zeta=0$  of the parameters in (\ref{BF}) gives:
\begin{equation}
B_{F}(\phi) = e^{ -\tau (\phi-\phi_{0})}   \,\, .
\end{equation}
In this case, if $ \tau \gta 1$ it is not possible to satisfy all the constraints at the same time.  Depending on the sign of $\tau$, the resulting $| \Delta \alpha / \alpha |$ becomes too large in the early or late universe.
For $\tau \ll 1$, instead, the coupling function becomes equivalent to the linear case: 
$\BF = e^{ -\tau (\phi-\phi_{0})} \simeq 1 -\tau (\phi-\phi_{0})$.

\subsection*{Linear and power--law coupling combined}
\begin{figure}[!htbp]
\begin{flushright}
\includegraphics[width=14cm]{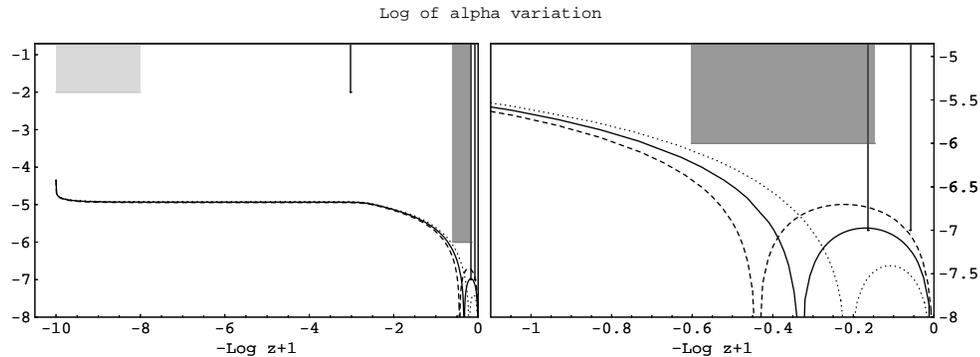}
\caption{The logarithm of $| \Delta \alpha / \alpha |$ is plotted as a function of Log$(z+1)$ for $B_{F}(\phi)= \left({\phi \over \phi_{0}}\right)^{\epsilon}(1-\gamma \, \epsilon \, (\phi-\phi_{0}))$ with $\epsilon = 2.4 \cdot 10^{-6}$,
$\gamma=2.2$ (solid line) and $\gamma=2.2 \pm 10\%$ (dashed and dotted respectively). 
On the right-hand side we zoom on $z\lta 10$. 
Only the curves not overlapping the grey areas are phenomenologically viable.}
\label{linear-power}
\end{flushright}
\end{figure}

Now let's consider two factors in (\ref{BF}) with $q=1$ and $\tau=0$. 
\begin{equation}
B_{F}(\phi)= \left({\phi \over \phi_{0}}\right)^{\epsilon} (1-\zeta \, (\phi-\phi_{0}))   \,\, .
\end{equation}
For an arbitrary choice of $\zeta$ and $\epsilon$, the resulting $\D \a$ is similar to the linear coupling or power--law coupling case, depending on which factor dominates.
It is instead interesting to consider the case $\zeta=\gamma \epsilon$ in which the two factors are of the same order of magnitude.
If $\gamma>0$, the two factors can contribute in an opposite way and it is easy to obtain $\Delta \alpha \simeq 0$  also at some time in the past. For example, with the choice $\epsilon=2.4 \cdot 10^{-6}$ and $\gamma=2.2$ we obtained the behavior plotted in Fig.~\ref{linear-power}, in which we have varied $\gamma$ of 10\%.
For the choice  $\epsilon = 2.4 \cdot 10^{-6}$, $\gamma=2.2$  the constraints on the violation of equivalence principle and the constraints derived from atomic clocks are automatically satisfied:
\begin{equation}
\eta \simeq 2 \cdot 10^{-20} \ll 10^{-13}
\qquad \qquad
\left | {\dot{\alpha} \over \alpha_{0}}\right |
=9 \cdot 10^{-17}
\ll 10^{-15} \mbox{ yr}^{-1}
\end{equation}
%
%
%

%
\subsection*{Power--law and exponential coupling combined}
Since, as already discussed, the exponential  coupling function case is equivalent to the linear one, this possibility falls within the previous example.

\subsection*{Polynomial and exponential combined}
\begin{figure}[htbp]
\begin{flushright}
\includegraphics[width=14cm]{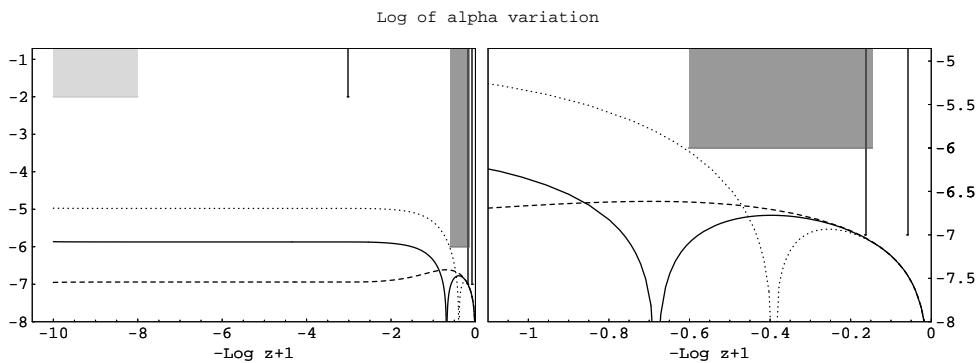}
\caption{The logarithm of $| \Delta \alpha / \alpha |$ is plotted as a function of Log$(z+1)$ for $B_{F}(\phi)=  (1-\zeta \, (\phi-\phi_{0})^6)\; e^{-\tau  (\phi-\phi_{0})}$ with $\tau=0.6 \cdot 10^{-6}$ and $\zeta=2 \cdot 10^{-4}$ (dotted line), $\zeta=3.2 \cdot 10^{-5}$ (solid line) and $\zeta=5 \cdot 10^{-6}$ (dashed line). 
On the right-hand side we zoom on $z\lta 10$. 
Only the curves not overlapping the grey areas are phenomenologically viable.}
\label{polinomial-exp}
\end{flushright}
\end{figure}

Now let's consider $\epsilon=0$ and $q\not = 1$. The case $q=1$ is not interesting since the two factors would be almost equivalent and the behavior  corresponding to the linear case.   Let's choose then, for example, $q=6$:
\begin{equation}
B_{F}(\phi)=  (1-\zeta \, (\phi-\phi_{0})^6)\; e^{-\tau  (\phi-\phi_{0})}
\end{equation}
For recent times ($z<1$) the exponential coupling $e^{-\tau  (\phi-\phi_{0})} \simeq 1- \tau  (\phi-\phi_{0})$ dominates, while in the past the two terms can be of the same order and, due to $q$ being even,  cancel at some time. This is shown in  Fig.~\ref{polinomial-exp}.
The constraints on the violation of equivalence principle and the constraints derived from atomic clocks are satisfied.

\section{Summary and Conclusions}
%
In this paper we have carried out a comprehensive study of the cosmological variation of the fine-structure constant $\a$ induced by the coupling of the electromagnetic field with a typical Quintessence scalar.
We have considered a variety of functional forms for the coupling function $\BF$, obtainable from a general expression (see Eq.~(\ref{BF})) depending on four parameters.

We have found that very different cosmological histories for $\D \a$ are possible, depending on which parameters are switched on. For example, we can produce a $\D \a$ which is well below the observational constraints in the early universe and just within the experimental limits in recent times (linear coupling case). But also the converse is possible, if we choose a polynomial coupling.
In particular, the behavior at small red-shift can be qualitatively very different depending on the model we choose: sharply decreasing in the polynomial coupling case or mildly decreasing with the power--law coupling.

By combining different functional forms, a notable feature emerged. In some cases it is possible that the scalar dynamics drives $\D\a$ to a zero at some time in the past, thus inverting the slope of its cosmological evolution. This happens for all the combined cases discussed here.

It is also worth remarking that in our parameter space span we found solutions with extremely reduced fine-tuning and still compatible with the available constraints. This is the case of the polynomial coupling with exponent $q \geq 15$ which lifts the fine-tuning of the coefficient $\zeta$ (usually constrained to be $\leq 10^{-6}$) to order 1.

It should also be emphasized that, while in the literature  the result by Webb et al. \cite{Webb,Murphy} is usually said to be incompatible with the Oklo limit \cite{Damour:1996zw}, we found that this is not always the case. For example, in the polynomial coupling case it is possible to obtain  several examples with a $\D \a$ which matches  the Quasars data and at the same time respects the Oklo bound.

\section*{Acknowledgments}
It is a pleasure to thank Antonio Masiero and Massimo Pietroni for useful discussions and suggestions.
FR is partially supported by the University of Padova fund for young researchers, research project No. CPDG037114.

%

\end{document}